\documentclass[runningheads]{llncs}

\usepackage[T1]{fontenc}
\usepackage{url}

\usepackage{caption}
\usepackage{array}
\usepackage{float}
\usepackage{placeins}
\usepackage{threeparttable}
\usepackage{graphicx}
\usepackage{multirow} 
\usepackage{amsmath}
\usepackage{tabularx}
\usepackage{booktabs}
\newcolumntype{L}[1]{>{\raggedright\arraybackslash}m{#1}} 
\newcolumntype{C}[1]{>{\centering \arraybackslash}m{#1}}

\usepackage[acronym]{glossaries}
\makeglossaries

\newacronym{xgbfs}{XGBFS}{XGBoost Feature Selection}
\newacronym{ids}{IDS}{Intrusion Detection System}
\newacronym{hids}{HIDS}{Host-based Intrusion Detection System}
\newacronym{ml}{ML}{Machine Learning}
\newacronym{dl}{DL}{Deep Learning}
\newacronym{pe}{PE}{Portable Executable}
\newacronym{rf}{RF}{Random Forest}
\newacronym{ember}{EMBER}{Elastic Malware Benchmark for Empowering Researchers}
\newacronym{lightgbm}{LightGBM}{Light Gradient-Boosting Machine}
\newacronym{auc}{AUC}{Area Under the Curve}
\newacronym{tpr}{TPR}{True Positive Rate}
\newacronym{fpr}{FPR}{False Positive Rate}
\newacronym{cnn}{CNN}{Convolutional Neural Networks}
\newacronym{sorel-20m}{SOREL-20M}{SOphos-REversingLabs 20 Million}
\newacronym{ffnn}{FFNN}{Feed-Forward Neural Network}
\newacronym{pca}{PCA}{Principal Component Analysis}
\newacronym{xgb}{XGBoost}{eXtreme Gradient Boosting}
\newacronym{et}{ET}{Extra Trees}
\newacronym{lstm}{LSTM}{Long-Short Term Memory}
\newacronym{api}{API}{Application Programming Interface}
\newacronym{lief}{LIEF}{Library to Instrument Executable Formats}
\newacronym{dll}{DLL}{Dynamic Link Library}
\newacronym{aws}{AWS}{Amazon Web Services}
\newacronym{lmdb}{LMDB}{Lightning Memory-Mapped Database}
\newacronym{json}{JSON}{JavaScript Object Notation}
\newacronym{jsonl}{JSONL}{JavaScript Object Notation Lines}
\newacronym{hdf5}{HDF5}{Hierarchical Data Format 5}
\newacronym{c2}{C2}{Command and Control}
\newacronym{iqr}{IQR}{InterQuartile Range}
\newacronym{ermds}{ERMDS}{Evaluating Robustness of Learning-based Malware Detection Systems}
\newacronym{enisa}{ENISA}{European Union Agency for Cybersecurity}
\newacronym{eu}{EU}{European Union}
\newacronym{raas}{RaaS}{Ransomware-as-a-Service}
\newacronym{eb}{EB}{EMBER + BODMAS}
\newacronym{ebr}{EBR}{EMBER + BODMAS + ERMDS}
\newacronym{nb}{NB}{Naïve Bayes}

\usepackage{biblatex}
\bibliography{refs}

\begin{document}

\title{Machine Learning Transferability for Malware Detection}

\author{César Vieira\orcidID{0009-0003-7906-1761} \and
João Vitorino\orcidID{0000-0002-4968-3653} \and
Eva Maia\orcidID{0000-0002-8075-531X} \and
Isabel Praça\orcidID{0000-0002-2519-9859}}

\authorrunning{César Vieira et al.}

\institute{GECAD, ISEP, Polytechnic of Porto, Rua Dr. António Bernardino de Almeida, 4249-015 Porto, Portugal, \email{\{csvva,jpmvo,egm,icp\}@isep.ipp.pt }}

\maketitle

\begin{abstract}
Malware continues to be a predominant operational risk for organizations, especially when obfuscation techniques are used to evade detection. 
Despite the ongoing efforts in the development of \acrfull{ml} detection approaches, there is still a lack of feature compatibility in public datasets. This limits generalization when facing distribution shifts, as well as transferability to different datasets.

This study evaluates the suitability of different data preprocessing approaches for the detection of \acrfull{pe} files with ML models. 
The preprocessing pipeline unifies EMBERv2 (2,381-dim) features datasets, trains paired models under two training setups: \acrfull{eb} and \acrfull{ebr}.
Regarding model evaluation, both \acrshort{eb} and \acrshort{ebr} models are tested against TRITIUM, INFERNO and SOREL-20M. ERMDS is also used for testing for the \acrshort{eb} setup.

The findings indicate that compact boosting static detectors are applicable for on-host use, but require a careful analysis of how PE obfuscation techniques affect the feature distributions of training datasets and during model inference.

\keywords{Malware detection, Static malware analysis, Obfuscation, Transferability, Machine learning}
\end{abstract}

\vspace*{-0.3cm}

\section{Introduction}

\vspace*{-0.2cm}

Malware is defined as software that intentionally compromises the confidentiality, integrity, or availability of information systems, thereby enabling attackers to engage in extortion, disruption and espionage activities \cite{NISTMalwareDef}. 
From 2023 to 2025, the \acrfull{enisa} identified the Public Administration as the most frequently targeted by malware within the \acrfull{eu}, with ransomware and data-related intrusions as the main attack vectors \cite{ENISA2023}, \cite{ENISA2024}.
These intrusions are indicative of trends that have been observed in the context of specialized attack ecosystems, which include \acrfull{raas} and access-brokered access \cite{ENISA2025}.

In order to prevent the execution of malicious programs, organizations can make use of both signature and anomaly-based approaches.
Signature-based detection approaches are capable of achieving high operational speeds and low \acrfull{fpr} by comparing the hash of a given binary with a database of known malware sample hashes or matching file content against a set of known byte or string patterns, such as YARA rules \cite{lockett2021}. 

On the other hand, anomaly-based detection approaches increasingly rely on \acrfull{ml} to learn a profile of benign host behavior or a discriminative boundary from static features, which requires hyperparameter optimization.
Methods include supervised, semi-supervised and unsupervised families, ranging from gradient-boosted trees and feed-forward networks to one-class and reconstruction-based approaches \cite{gaber2024}. 

However, despite the growing usage of \acrshort{ml} in anomaly-based approaches, their application to real-world \acrfull{pe} files still faces challenges.
The limited compatibility and standardization of the utilized features across public datasets hinders the reproducibility and transferability of detection pipelines \cite{raff2020}.
Moreover, models trained on one dataset may generalize poorly to others due to distributional mismatch and evolving attacker tooling, which may cause concept drift over time, requiring periodic model retraining and recalibration of feature distributions and data preprocessing approaches \cite{TranscendingTranscend}. 

This study evaluates different data preprocessing approaches for the detection of \acrshort{pe} files with ML models.
A static detection pipeline is built on \acrshort{ember}-v2 features with two training setups: \acrlong{eb} and \acrlong{ebr}.
\acrfull{pca} or \acrfull{xgbfs} are then applied to the unified datasets, producing feature vectors of 128/256/384 dimensions where FLAML tuned model pairs (\acrshort{lightgbm}, \acrshort{xgb}, \acrlong{et}, \acrlong{rf}) are then trained against the aforementioned reduced vector dimensions.
Evaluation covers in-distribution test splits and cross-dataset inference on \acrshort{sorel-20m}, TRITIUM, INFERNO (and \acrshort{ermds} when withheld), reporting F1, \acrshort{auc} and \acrshort{tpr}@1\%/0.1\% \acrshort{fpr} to assess generalization, obfuscation robustness and drift sensitivity.

This paper is divided into multiple sections.
Section \ref{sec:Related Works} reviews existing work regarding malware detection. 
Section \ref{sec:Methodology} describes the workflow of the study. 
Section \ref{sec:Results} presents and discusses the obtained results.
Finally, section \ref{sec:Conclusion} summarizes the main findings and outlines future work on generalization and concept drift.

\vspace*{-0.3cm}

\section{Related Work}
\label{sec:Related Works}

\vspace*{-0.2cm}

To choose the most relevant datasets and data preprocessing approaches, it is important to analyze recent literature regarding static Windows malware detection. Table \ref{Static malware detection} provides a summary of relevant scientific papers, covering the key milestones and more recent developments.

Historically, signature-based detection on Windows has relied on deterministic content rules that require minimal computation. 
However, such systems are vulnerable to semantics-preserving changes, packing, polymorphism and other forms of obfuscation, which limit the coverage of novel and zero-day malware. 
Motivated by these limitations, Schultz et al. \cite{schultz2001} and Kolter \& Maloof \cite{kolter2004} proposed approaches that relied on the extraction of features such as imported functions, strings and byte n-grams, which were then processed by \acrshort{ml} models like \acrfull{nb} and boosted decision trees.

Subsequently, Saxe \& Berlin \cite{saxe2015} replaced manual feature selection with automatic representations derived from two-dimensional byte-entropy histograms of entire binaries. 
This approach captured spatial correlations and entropy variations within executable content, improving scalability, robustness to obfuscation and higher classification accuracy compared to manually engineered n-gram features, marking an early transition toward deep-learning static malware analysis.

\vspace{-0.85cm}

\begin{table}[!htbp]
\caption{Overview of static malware detection approaches.}
\small
\label{Static malware detection}
\centering
\newcommand{\PaperW}{0.30\textwidth}
\newcommand{\DataW}{0.20\textwidth}
\newcommand{\ApproachW}{0.28\textwidth}
\newcommand{\MetricsW}{0.16\textwidth}

\renewcommand{\arraystretch}{1.11}
\resizebox{\columnwidth}{!}{%
\begin{tabular}{r@{\quad}C{\PaperW}L{\DataW}L{\ApproachW}L{\MetricsW}}
\hline
\multicolumn{1}{l}{\rule{0pt}{12pt}\textbf{Year}} &
\multicolumn{1}{c}{\textbf{Paper}} &
\multicolumn{1}{c}{\textbf{Data}} &
\multicolumn{1}{c}{\textbf{Approach}} &
\multicolumn{1}{c}{\textbf{Metrics}}\\[2pt]
\hline \rule{0pt}{12pt}

2001 & \textit{Schultz et al.}~\cite{schultz2001}
     & Proprietary \acrshort{pe}
     & Imports/strings/byte n-grams $\to$ NB, boosted trees
     & \acrshort{tpr} 0.97; \acrshort{fpr} 0.04; Acc 97.1\% \\

2004 & \textit{Kolter \& Maloof}~\cite{kolter2004}
     & Proprietary \acrshort{pe}
     & Byte n-grams $\to$ Boosted J48
     & \acrshort{auc} 0.9958 \\

2015 & \textit{Wang et al.}~\cite{wang2015}
     & BIG 2015
     & Ensemble (RF \& GBM) on byte/ASM features
     & 0.0028 Log Loss \\

2015 & \textit{Saxe \& Berlin}~\cite{saxe2015}
     & Proprietary \acrshort{pe}
     & \acrshort{cnn} on 2D byte–entropy images
     & $\uparrow$ vs.\ engineered n-grams \\

2018 & \textit{Raff et al.}~\cite{raff2018}
     & Windows \acrshort{pe} corpora
     & 1D \acrshort{cnn} on raw bytes
     & Acc 0.940 \& \acrshort{auc} 0.981 \\

2018 & \textit{Anderson \& Roth}~\cite{anderson2018}
     & \acrshort{ember} 2017/2018
     & LGBM
     & \acrshort{auc} $\approx$ 0.999; 92.9\% \acrshort{tpr}@0.1\% \acrshort{fpr} \\

2019 & \textit{Rudd et al.}~\cite{rudd2019}
     & Proprietary \acrshort{pe}
     & \acrshort{ffnn} with auxiliary losses
     & $\Delta$FNR $-42.6\%$@$10^{-3}$; $-53.8\%$@$10^{-5}$ \\

2020 & \textit{Harang \& Rudd}~\cite{harang2020}
     & \acrshort{sorel-20m}
     & \acrshort{lightgbm}; multi-task \acrshort{ffnn} (ALOHA-style)
     & \acrshort{auc} $\approx$ 0.998; \acrshort{tpr} $>$90\%@0.1\% \acrshort{fpr} \\

2021 & \textit{Rahali et al.}~\cite{rahali2021}
     & \acrshort{ember}
     & Cost-sensitive LGBM (custom logistic loss)
     & $\downarrow$ error vs.\ LGBM \\

2023 & \textit{Austin et al.}~\cite{austin2023}
     & \acrshort{ember}, \acrshort{sorel-20m}
     & AutoML-tuned \acrshort{ffnn} (static \& online settings)
     & 0.956 \acrshort{tpr}@0.1\% (\acrshort{ember}); 0.965 (\acrshort{sorel-20m}) \\

2023 & \textit{Manikandaraja et al.}~\cite{manikandaraja2023}
     & TRITIUM, INFERNO
     & \acrshort{ember}, BODMAS \& SOREL classifiers
     & TRITIUM F1/\acrshort{auc}: 0.904/0.913; INFERNO: 0.959/0.961  \\

2025 & \textit{Choudhary et al.}~\cite{choudhary2025}
     & \acrshort{ember}, BODMAS, \acrshort{ermds}
     & Dim.\ reduction (\acrshort{pca}/\acrshort{xgb}) + two-instance soft-vote (\acrshort{rf}/ \acrshort{et}/ \acrshort{lightgbm}/ \acrshort{xgb})
     & F1 $\approx$ 0.9773; \acrshort{auc} $\approx$ 0.9958 \\[2pt]

\hline
\end{tabular}
}
\end{table}
\FloatBarrier

The earliest major benchmark for static malware classification was Microsoft BIG 2015 \cite{ronen2015}, which provided over 500 GB of labeled Windows \acrlong{pe} binaries grouped into nine malware families. 
The competition instigated the first comparative studies between traditional feature-engineered models and emerging deep-learning methods for family classification.
The solution that was found to be most effective by Wang et al. \cite{wang2015} used an ensemble of \acrfull{rf} and Gradient Boosting on engineered byte and assembly-level statistics.

The publication of \acrfull{ember} \cite{anderson2018} contributed to the advancement of static malware classification by introducing a fully open and standardized dataset for static malware detection.
The dataset includes 1.1 million \acrshort{pe} files, encoded into 2381-dimensional feature vectors, capturing \acrshort{pe} header fields, section entropies, imports and string statistics through feature hashing.
The \acrfull{lightgbm} baseline achieved an \acrfull{auc} of 0.999, with a 92.9\% \acrshort{tpr}@0.1\% \acrshort{fpr}, surpassing the byte-sequence \acrfull{cnn} MalConv model, which achieved an \acrshort{auc} of 0.998.
The authors further updated the dataset with the introduction of the 2018 version, which included samples from 2018 and expanded the representation to a 2381-feature vector.

Further work also saw improvements on the baseline model, where Rahali et al. \cite{rahali2021} proposed an adaptation of \acrshort{lightgbm} incorporating a custom logistic loss function in order to account for the cost of misclassifying samples.
When evaluated against the standard \acrshort{lightgbm}, their approach achieved a lower classification error rate and demonstrated more stable convergence behaviour.

To address the limitations in scope and timeframe imposed by \acrshort{ember}, \acrfull{sorel-20m} \cite{harang2020} introduced a dataset containing approximately 20 million \acrshort{pe} samples. 
The latter takes advantage of the \acrshort{ember}-v2 feature standard and uses temporal splits for training, validation and testing.
The models applied to assess the dataset were \acrshort{lightgbm} and a multi-task \acrfull{ffnn} inspired by the ALOHA approach \cite{rudd2019}.
The results of the experiment yielded an $\approx$ 0.998 \acrshort{auc} and \acrshort{tpr}@0.1\% \acrshort{fpr} above 90\%.
Subsequent studies, including Austin B. et al. \cite{austin2023}, have employed AutoML-driven hyperparameter optimization in the context of \acrshort{ffnn} architectures that have been trained on both \acrshort{ember} and \acrshort{sorel-20m}.  
The study reported improvements in the low \acrshort{tpr} regime across both datasets, reaching a \acrshort{tpr}\@0.1\%\acrshort{fpr} of 0.956 on \acrshort{ember} and 0.965 on \acrshort{sorel-20m}.

Recent research highlights the persistent challenge of concept drift, defined as the evolution of malware features at a rate faster than static detectors can adapt.
The Rapidrift framework, proposed by Manikandaraja et al. \cite{manikandaraja2023}, is based on this premise and introduces a methodology capable of analyzing concept drift at a more granular level, quantifying the degradation of model performance across temporal partitions and adversarial shifts. In the same study, the authors released two new datasets, TRITIUM and INFERNO, designed to capture distinct aspects of real-world malware evolution. TRITIUM is a product of naturally occurring Windows \acrshort{pe} samples collected from operational environments, whereas INFERNO was constructed using popular cybersecurity tools that allow the creation and modification of malicious payloads. 
Furthermore, Choudhary A. et al. \cite{choudhary2025} explored ensemble and gradient boosting classifiers along dimensionality-reduction techniques such as \acrlong{pca} and \acrlong{xgbfs} over  \acrshort{ember}, BODMAS and \acrfull{ermds}. The study obtained an F1 score of $\approx$ 0.9773 and $\approx$ 0.9958 \acrshort{auc} score, while simultaneously reducing vector dimensionality by nearly 85\%.

In summary, an analysis of current literature reveals an inability to deliver detectors capable of generalizing across heterogeneous systems, feature pipelines and temporal shifts, with obfuscation often degrading performance outside the training domain.
The analysis of existing studies reveals a predominant tendency to evaluate primarily on self-curated datasets and feature variants, which imposes limitations on the comparison of results and complicates cross-dataset generalization and robustness to packing, polymorphism and concept drift.

\vspace*{-0.3cm}

\section{Methodology}
\label{sec:Methodology}

\vspace*{-0.2cm}

This section describes the work pipeline for both approaches and the datasets used throughout the study. 
The performed experiments included two different training datasets, \acrshort{eb} and \acrshort{ebr}, and two different feature reduction processes, \acrshort{pca} and \acrshort{xgbfs}. These were used to train \acrshort{lightgbm}, \acrshort{xgb}, \acrlong{et} and \acrlong{rf} model pairs, with hyperparameters tuned via FLAML and selected on the corresponding validation split of each unified dataset.
The experiments were performed on a Windows Desktop with an Intel i7-12700KF, a NVIDIA 5070 TI and 32 GB of RAM, occupying $\approx$ 200 GB of disk space. 

\vspace*{-0.3cm}
\subsection{Datasets}
\label{subsec:datasets}

This study used six open Windows \acrshort{pe} datasets, that capture both real-world and adversarial conditions. Table \ref{tab:datasets_overview} provides a quick overview for each.

\vspace*{-0.15cm}
\begin{table}[!htbp]
\caption{Datasets characteristics}
\label{tab:datasets_overview}
\centering
\resizebox{\columnwidth}{!}{%
\begin{tabular}{L{0.18\textwidth} L{0.33\textwidth} L{0.53\textwidth}}
\hline
\textbf{Dataset} & \textbf{Size (1 / 0 / -1)} & \textbf{Notes} \\
\hline
\acrshort{ember}-2018 &
1.1 M (400k / 400k / 300k) &
Files scanned in or before 2018 \\

\acrshort{sorel-20m} &
$\approx$ 20 M (10M / 10M) &
1 Jan 2017 - 10 Apr 2019; temporal splits. \\

BODMAS &
$\approx$ 134 K (57,293 / 77,142) &
29 Aug 2019–30 Sept 2020; 581 families. \\

\acrshort{ermds}* &
$\approx$ 106 K (30,229 / 76,662) &
Jan - Dec 2022; Binary/source/packing obfuscations. \\

TRITIUM &
$\approx$ 37 K (19,712 / 17,785)&
Naturally occurring 2022 threat samples. \\

INFERNO &
2,864 (1,432 / 1,432) &
Red-team/custom \acrshort{c2} malware. \\
\hline
\end{tabular}%
}
\begin{tablenotes}
    \item * - Number of samples after unifying the subsets and de duplicating samples 
\end{tablenotes}
\end{table}
\FloatBarrier

\paragraph{\textbf{\acrshort{ember}-2018}.}
Anderson \& Roth \cite{anderson2018} released the \acrshort{ember} dataset, which is a large-scale, open benchmark designed for the static classification of malware on Windows \acrshort{pe} files. Each binary is parsed into structured \acrfull{jsonl} records using the \acrfull{lief} framework. This yields eight major feature groups, more concretely, (1) general file information, (2) header information, (3) optional-header data directories, (4) imported functions, (5) exported functions, (6) section metadata, (7) byte-entropy histograms and (8) string-statistics histograms.

Numerical features, such as header sizes, section entropies and file offsets, are retained as scalars. Meanwhile, variable-length textual fields, such as import and section names and \acrfull{dll}s, are transformed using a feature-hashing trick to produce a fixed-length vector. Each vectorized feature yields a 2,381-dimensional numerical representation vector per sample. Furthermore, the code used to extract these features from other \acrshort{pe} files is also used in the datasets included in this study and available in their GitHub Repository\footnote{https://github.com/elastic/ember}.

\paragraph{\textbf{\acrshort{sorel-20m}}.}
Harang \& Rudd \cite{harang2020} introduced \acrshort{sorel-20m}, a large-scale benchmark that extends the \acrshort{ember} paradigm to real-world, temporally structured malware detection. The corpus comprises approximately 20 million Windows \acrshort{pe} samples from late 2017 to 2020, with binaries disarmed to preserve structural metadata while enabling safe feature extraction.

It is distributed via the \acrfull{aws} Open Data Registry and supporting utilities are available in the official GitHub repository\footnote{https://github.com/sophos/SOREL-20M}. This includes pre-computed \acrshort{ember}-v2 feature vectors alongside auxiliary metadata, such as vendor detection counts and behavioural tags, as well as per-sample SHA-256 and first-seen timestamps. These are provided in \acrfull{lmdb} files and a companion SQLite3 metadata database.

\paragraph{\textbf{BODMAS}.}
Yang et al. \cite{bodmas} released BODMAS\footnote{https://github.com/whyisyoung/BODMAS} to support both temporal analysis and family-aware evaluation of static Windows \acrshort{pe} detectors. The corpus contains $\approx$ 134,000 samples collected from August 2019 to September 2020, with 581 family labels. The aim was to complement existing benchmark datasets by providing more recent, timestamped samples and detailed family labels for studying concept drift, family attribution and cross-dataset generalization.

The public release provides pre-extracted feature vectors and metadata. Specifically, the feature vectors are distributed as a NumPy archive (\textit{bodmas.npz}), while \textit{bodmas\_metadata.csv} provides the SHA-256 hash and timestamp for each sample and \textit{bodmas\_malware\_category.csv} the family label for malware samples.

\paragraph{\textbf{\acrshort{ermds}}.}
Jia et al. \cite{ermds} introduced \acrshort{ermds} to address learning-based malware detection systems on detecting obfuscated malware. The dataset is divided into three subsets for specific obfuscation spaces: binary-level, source code and packer obfuscation. The combined dataset, labelled \acrshort{ermds}-X, contains $\approx$ 106,000 samples and uses the same \acrshort{ember}-v2 feature processing pipeline.

The release of \acrshort{ermds}\footnote{https://github.com/lcjia94/ERMDS} includes pre-extracted features in \acrfull{json} format for each subset and scripts to expand the dataset and evaluate MalConv and \acrshort{lightgbm} \acrshort{ember} models.

\paragraph{\textbf{TRITIUM}.}
Manikandaraja et al. \cite{manikandaraja2023} introduced TRITIUM and INFERNO as part of the Rapidrift framework, which were developed to evaluate concept drift and model generalization in \acrshort{ml}-based malware detection. 
TRITIUM provides $\approx$ 37,000 Windows \acrshort{pe} samples, which are stored in two separate \acrfull{hdf5} files: \textit{tritium.h5} ($\approx$ 23,000) and \textit{tritium\_unseen.h5} ($\approx$ 14,000), extracted from naturally occurring threats collected in 2022. 
The dataset is publicly available through its GitHub repository\footnote{https://github.com/4dsec/tritium}.

\paragraph{\textbf{INFERNO}.}
INFERNO\footnote{https://github.com/4dsec/inferno}, released concurrently with TRITIUM, consists of around 1,430 malicious samples derived from red team and custom \acrfull{c2} malware. It was designed to emulate evasive and adversarial behaviors not present in earlier datasets, with the aim of providing a challenging benchmark for evaluating robustness and zero-day generalization.

\vspace{-0.3cm}

\subsection{Data Processing}

\begin{figure}[ht]
\centering
\includegraphics[width=\textwidth]{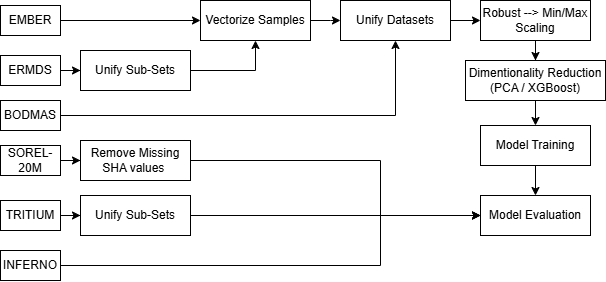}
\caption{Data preprocessing and model evaluation workflow.}
\label{fig:Workflow}
\end{figure}

The design and implementation of the presented processing pipeline closely follow the methodology proposed by Choudhary et al. \cite{choudhary2025}, with some adaptations regarding dataset usage. 
As outlined in Section \ref{subsec:datasets}, all datasets are structured in accordance with the \acrshort{ember}-v2 feature standard \cite{anderson2018}. 
The workflow, illustrated in Figure \ref{fig:Workflow}, begins with the dataset-specific preprocessing steps, which include vectorization, the removal of missing samples and the unification of sub-sets. 

Subsequently, a unified training dataset consisting of the \acrshort{eb} is produced and model training splits the combined training and validation pools into two equal partitions.
The remaining datasets are intended exclusively for the purpose of evaluating the models robustness against obfuscated specific attacks and large data in post-training inference evaluation.

Following the consolidation of the dataset, Robust Scaling is applied in order to mitigate the influence of outliers in the data. 
This scaling method centers features around their median and scales by the \acrfull{iqr}, providing resilience against the potentially distorting effect of extreme values. This robust normalization is particularly suitable when datasets contain skewed distributions, as it preserves the primary data structure while stabilizing the learning process \cite{deAmorim2023}.
Afterwards, MinMax Scaling is applied, rescaling each feature to a fixed interval by linearly transforming all values into an interval between 0 and 1. 
This guarantees consistent input ranges, facilitating stable optimization and fair feature comparison in downstream models. 
All scaling parameters are determined exclusively from the training set, ensuring strict data segregation across validation and test phases.

Consequently, the feature vectors undergo dimensionality reduction using supervised feature selection via \acrshort{xgbfs}, where features are ranked according to their importance gain and unsupervised \acrshort{pca}, which projects data into orthogonal components, thus preserving maximal variance. 
For each technique it was produced reduced feature subsets of 128, 256 and 384 dimensions.

\vspace*{-0.3cm}

\subsection{Model Training}
For each tree classifier used, namely \acrshort{xgb}, \acrshort{lightgbm}, \acrlong{et} and \acrlong{rf}, two independent instances were trained, one per partition and tune each instance with FLAML \cite{flaml} to optimize its hyperparameters on its own partition. 
This yields a pair of tuned models per classifier that have seen separate training data and undergone separate searches.

At inference the two instances of a given classifier were combined by weighted soft voting. 
Let $x$ be a sample while $p_{1}(x)$ and $p_{2}(x)$ are the probability outputs for each instance, the final score is calculated as in Equation \ref{eq:final-classification}:

\begin{equation}
\label{eq:final-classification}
 \hat{p}(x;w)=w \cdot p_{1}(x) + (1-w) \cdot p_{2}(x),w \in [0,1]   
\end{equation}
where $w$ increments in steps of 0.1. 
The weight pairs that achieve the highest validation performance are kept for testing and all robustness evaluations. 

For each reduced feature set, F1-score, \acrshort{auc} and \acrshort{tpr}@0.1\% and 1\% \acrshort{fpr} were used to assess models efficiency. 
This choice for evaluation metrics is justified since F1-Score serves as an harmonic mean of precision and recall at a chosen decision threshold (in the case of this study, best\_f1).
\acrshort{auc} helps understand if the model is actually learning from the features, while \acrshort{tpr} at a threshold allows for an estimation of the model performance in real environments, since malware only constitutes a small percentage of sample in these cases. 

\vspace*{-0.3cm}

\section{Results and Discussion}
\label{sec:Results}

\vspace*{-0.2cm}

Since this study is focused on dataset generalization, the evaluation was performed not only with the testing sets of \acrshort{eb} and \acrshort{ebr}, but also with the TRITIUM, INFERNO and \acrshort{ermds} datasets. 
The transferability of the \acrshort{ml} models is assessed using multiple evaluation metrics, such as F1-Score, \acrshort{auc} and \acrshort{tpr}@1\% and 0.1\% \acrshort{fpr}. \acrshort{sorel-20m}, on the other hand, was only tested using the test split due to limitations in computational resources.

Regarding the \acrshort{eb} approach outlined in Table \ref{tab:EMBER_BODMAS}, \acrshort{xgbfs} consistently outperformed \acrshort{pca} at matched dimensionality. 
Performance improved with increasing dimensionality, as the 384-dim setting retained more informative signals while effectively removing noise, whereas 128/256 dims discarded weak yet important cues. 
At 384 dims, \acrshort{lightgbm} reached 99.84\% \acrshort{auc}, indicating strong class separation. 
Operating-point results confirm this, where at 1\% \acrshort{fpr} the detector identifies 97.50\% of malware and at 0.1\% \acrshort{fpr}  91.25\%, suitable for enterprise use where false alarms are unfavorable. 
The F1 of 98.27\% indicates well-balanced precision and recall.

\vspace*{-0.15cm}

\begin{table}[!htbp]
\small
\centering
\caption{Dimensionality Reduction Results on \acrshort{ember} \& BODMAS}
\label{tab:EMBER_BODMAS}
\begin{tabular}{|c|c|c|c|c|c|c|}
\hline
\textbf{red.} & 
\textbf{dim.} & 
\textbf{est.} & 
\textbf{F1 (\%)} & 
\textbf{\acrshort{auc} (\%)} &
\textbf{1\%\acrshort{fpr} (\%)} &
\textbf{0.1\%\acrshort{fpr} (\%)} 
\\
\hline
\multirow{3}{*}{\acrshort{pca}}    & 128 & LGBM & 97.16 & 99.64 & 94.50 & 86.98 \\
                        \cline{2-7}
                        & 256 & XGB & 97.20 & 99.66 & 94.87 & 86.70 \\
                        \cline{2-7}
                        & \textbf{384} & \textbf{XGB} & \textbf{97.27} & \textbf{99.66} & \textbf{94.26} & \textbf{84.04} \\
\hline
\multirow{3}{*}{XGBFS}  & 128 & LGBM & 97.71 & 99.75 & 95.69 & 86.77 \\
                        \cline{2-7}
                        & 256 & LGBM & 98.26 & 99.84 & 97.48 & 91.43 \\
                        \cline{2-7}
                        & \textbf{384} & \textbf{LGBM} & \textbf{98.27} & \textbf{99.84} & \textbf{97.50} & \textbf{91.25} \\
\hline
\end{tabular}
\end{table}
\FloatBarrier

The integration of \acrshort{ermds} into the training process yielded consistent outcomes, as depicted in Table \ref{tab:EMBER_BODMAS_ERMDS}. 
A slight decrease in performance at low \acrshort{fpr} was observed in comparison to the performance of the \acrshort{eb} approach. 
This decline can be explained by the fact that the samples contributed by \acrshort{ermds} are obfuscation-heavy, resulting in a shift in the feature distribution and an increase in the intra-class variance. 
This leads to a reduction in the margin between benign and malicious samples, as the feature vectors become more spread out within the same class.
The optimal model, \acrshort{lightgbm}, obtained 98.12\% F1, 99.82\% \acrshort{auc}, 96.90\% \acrshort{tpr}@1\%\acrshort{fpr} and 89.61\% \acrshort{tpr}@0.1\%\acrshort{fpr}.
Across both regimes, \acrshort{lightgbm} excelled compared to the other classifiers, while \acrshort{pca} did not match the performance of \acrshort{xgbfs} at the same vector sizes.  

\vspace*{-0.15cm}

\begin{table}[!htbp]
\small
\centering
\caption{Dimensionality Reduction Results on \acrshort{ember}, BODMAS and \acrshort{ermds}}
\label{tab:EMBER_BODMAS_ERMDS}
\begin{tabular}{|c|c|c|c|c|c|c|c|c|c|}
\hline
\textbf{red.} & 
\textbf{dim.} & 
\textbf{est.} & 
\textbf{F1 (\%)} & 
\textbf{\acrshort{auc} (\%)} & 
\textbf{1\%\acrshort{fpr} (\%)} & 
\textbf{0.1\%\acrshort{fpr} (\%)} 
\\
\hline
\multirow{3}{*}{\acrshort{pca}}     & 128 & ET & 96.50 & 99.44 & 91.27 & 74.01 \\
                                    \cline{2-7}
                                    & 256 & XGB & 96.42 & 99.47 & 91.05 & 78.94 \\
                                    \cline{2-7}
                                    & \textbf{384} & \textbf{XGB} & \textbf{96.51} 
                                    & \textbf{99.49} & \textbf{91.21} & \textbf{79.59} \\
\hline
\multirow{3}{*}{XGBFS}              & 128 & ET & 97.73 & 99.76 & 95.56 & 87.69 \\
                                    \cline{2-7}
                                    & 256 & LGBM & 98.00 & 99.80 & 96.57 & 88.82 \\
                                    \cline{2-7}
                                    & \textbf{384}  & \textbf{LGBM} & \textbf{98.12}                & \textbf{99.82} & \textbf{96.90} & \textbf{89.61} \\
\hline
\end{tabular}
\end{table}
\FloatBarrier
Additionally, the 24 trained model pairs from each approach were evaluated with the rest of test datasets.
The results in Table \ref{tab:infer_results} displays the best performing model for each dataset as well as the results from the best performing model for the corresponding training set, for brevity. 
All the best performing models used \acrshort{xgbfs} as it's dimensionality reduction algorithm.

\vspace*{-0.15cm}
\begin{table}[!htbp]
\small
\centering
\caption{Model pair results against other \acrshort{ember}-v2 datasets}
\label{tab:infer_results}
\begin{tabular}{|c|c|c|c|c|c|c|}
\hline
\textbf{Dataset} &
\textbf{Dim.} &
\textbf{Est.} & 
\textbf{F1 (\%)} & 
\textbf{\acrshort{auc} (\%)} & 
\textbf{1\%\acrshort{fpr} (\%)} &
\textbf{.1\%\acrshort{fpr} (\%)} 
\\
\hline
\multirow{4}{*}{INFERNO}    & 128 & $\text{RF}^{(1)}$ & 94.54 & 98.51 & 84.66 & 46.38 \\
                            \cline{2-7}
                            & 384 & $\text{LGBM}^{(1)}$ & 92.30 & 97.72 & 86.47 & 81.91 \\
                            \cline{2-7}
                            & \textbf{384} & \textbf{$\text{XGB}^{(2)}$} & \textbf{95.43} & \textbf{98.90} & \textbf{86.98} & \textbf{85.75} \\
                            \cline{2-7}
                            & 384 & $\text{LGBM}^{(2)}$ & 93.04 & 98.40 & 86.18 & 45.15 \\
\hline
\multirow{3}{*}{TRITIUM}    & \textbf{384} & \textbf{$\text{LGBM}^{(1)}$} & \textbf{98.22} & \textbf{99.79} & \textbf{96.28} & \textbf{81.10} \\
                            \cline{2-7}
                            & 384 & $\text{XGB}^{(2)}$ & 98.62 & 99.83 & 97.13 & 65.89 \\
                            \cline{2-7}
                            & 384 & $\text{LGBM}^{(2)}$ & 98.07 & 99.76 & 95.96 & 77.54 \\
\hline
\multirow{4}{*}{\acrshort{sorel-20m}} & \textbf{384} & \textbf{$\text{ET}^{(1)}$} & \textbf{85.54} & \textbf{95.37} & \textbf{54.64} & \textbf{30.09} \\
                            \cline{2-7}
                            & 384 & $\text{LGBM}^{(1)}$ & 83.50 & 94.92 & 35.46 & 14.78 \\
                            \cline{2-7}
                            & 128/PCA & $\text{XGB}^{(2)}$ & 69.68 & 84.76 & 25.40 & 09.44 \\
                            \cline{2-7}
                            & 384 & $\text{LGBM}^{(2)}$ & 56.73 & 65.42 & 06.04 & 00.19 \\
\hline
\acrshort{ermds} & 384 & $\text{LGBM}^{(1)}$ & 83.73 & 55.03 & 09.36 & 02.97 \\
\hline
\end{tabular}
\begin{tablenotes}
    \item 1 - Model trained with \acrshort{eb}; 2 - Model trained with \acrshort{ebr}
\end{tablenotes}
\end{table}
\FloatBarrier

The findings presented in Table \ref{tab:infer_results} demonstrate that both INFERNO and TRITIUM consistently achieve high F1 and AUC scores, along with competitive low \acrshort{fpr} operating points, suggesting their capacity for broad-scale generalization across smaller external datasets. 
Conversely, both \acrshort{sorel-20m} and \acrshort{ermds} exhibit significant deterioration across all metrics, with the most substantial declines occurring at 1\% and 0.1\% \acrshort{fpr}, reflecting their sensitivity to both domain and temporal shifts, as well as to obfuscation. 
The performance of the system is dependent on the specific estimators and training methods employed. 
In general, the combination of boosting classifiers and \acrshort{xgbfs} tends to maintain a more stable low \acrshort{fpr} behavior, however the system remains susceptible to severe drift.

\vspace*{-0.3cm}

\section{Conclusions}
\label{sec:Conclusion}

\vspace*{-0.2cm}

The present study investigated host-based detection on Windows operating systems using static \acrshort{ember}-v2 features and two training regimes, namely \acrshort{eb} and \acrshort{ebr}. 
The pipeline made use of robust and min–max scaling, reduced dimensionality to 128, 256 and 384 using \acrshort{pca} or \acrshort{xgbfs}, trained model pairs for four tree estimators with FLAML-tuned hyperparameters and evaluated model performance with F1, \acrshort{auc} and \acrshort{tpr} at 1\% and 0.1\% \acrshort{fpr} on a unified test splits and other \acrshort{ember}-v2 datasets.

The results of this study indicate that boosting-based estimators are the most reliable, especially when paired with \acrshort{xgbfs} at 384 dimensions, which consistently outperforms \acrshort{pca} at matched sizes. 
Model generalization remained sensitive to the training mixture.
Specifically, \acrshort{eb}  models performance degraded substantially when tested against \acrshort{ermds}, whereas including \acrshort{ermds} in the training set reduced generalization to \acrshort{sorel-20m} relative to \acrshort{eb}. 
This pattern is consistent with a distributional mismatch between \acrshort{ermds} and the remaining \acrshort{ember}-v2 datasets. 
The \acrshort{eb} configuration was particularly vulnerable to obfuscation-heavy samples when \acrshort{ermds} was treated as external data, which motivates further analysis of how obfuscation affects feature distributions.

Future work will include additional models proposed in the literature which include additional \acrshort{ml} models and more complex deep learning models in order to assess robustness to obfuscation and concept drift, focusing on cross-dataset generalization. 
Overall, it is important to continue the research efforts into the analysis of how feature representation, training data composition and design choices affect model generalization, especially at low \acrshort{fpr}.

\subsubsection*{Acknowledgments.}
This work has been supported bythe PC2phish project, which has received funding from FCT with Refª:2024.07648.IACDC. Furthermore, this work also received funding from the project UID/00760/2025.

\begingroup
\small
\printbibliography 
\endgroup
\end{document}